\documentstyle[preprint,aps]{revtex}
\begin{document}

\draft


\title{The non-Markovian stochastic Schr\"odinger equation for open systems}

\author{Lajos Di\'osi$^{1,2,}$\footnote{e-mail: diosi@rmki.kfki.hu}
 and Walter T. Strunz$^{1,3,}$\footnote{e-mail: walter.strunz@uni-essen.de}}

\address{$^1$Department of Physics, Queen Mary and Westfield College, University
         of London,\\ Mile End Road, London E1 4NS, UK\\
         $^2$ KFKI Research Institute for Particle and Nuclear Physics, 
 H-1525 Budapest 114, PO Box 49, Hungary\\
         $^3$Fachbereich Physik, Universit\"at GH Essen, D-45117 Essen, Germany}

\date{June 18, 1997}

\maketitle

\begin{abstract}
We present the non-Markovian generalization of the widely used stochastic
Schr\"odinger equation. Our result allows to 
describe open quantum 
systems in terms of stochastic state vectors rather than density operators,
without Markov approximation. 
Moreover, it unifies two recent independent attempts towards a
stochastic description of non-Markovian open systems, based on
path integrals on the one hand and coherent states on the other.
The latter approach utilizes the analytical properties of coherent states
and enables a microscopic interpretation of the stochastic states.
The alternative first approach is based on the general description of
open systems using path integrals as originated by Feynman and Vernon.
\end{abstract}


In the last few years the description of open quantum systems in terms
of stochastic Schr\"odinger equations
has received remarkable attention. They are now widely used 
in different fields (measurement theory, quantum optics, quantum chaos,
solid states
\cite{Diosi95,GisPer,Carmic94,GoeGra94,PleKni97,SpiRal94,Viola97}),
wherever quantum irreversibility matters. They do not only serve as a fruitful 
theoretical concept but also as a practical method for computations in the
form of quantum trajectories.
Up to now, however, the Markov approximation was believed to be
essential for a stochastic description \cite{KleSha95}. 
For systems where
non-Markovian effects are inevitable, as for non-equilibrium relativistic
fields, especially in quantum cosmology \cite{cosmos} or solid state
physics \cite{Graber88Weiss93,solid}, an advantageous stochastic pure
state description was missing.
This Letter
presents an exact non-Markovian stochastic Schr\"odinger equation.

Traditionally, open systems are described by the reduced density 
operator
$\hat\rho_{sys}(t) = \mbox{Tr}_{env}\left(\hat\rho_{tot}(t)\right),$
obtained from the total density operator by tracing over the environmental
degrees of freedom.
In Markov approximation, it is well known that the system 
dynamics
can either be described by a master equation for the reduced density 
operator $\hat\rho_{sys}(t)$,
or alternatively, by a stochastic Schr\"odinger equation for state vectors
$|\psi_Z(t)\rangle$
\cite{Diosi95,GisPer,Carmic94,GoeGra94,PleKni97,SpiRal94,Viola97}.
In this latter approach, 
the reduced density operator is recovered as the ensemble average
over these stochastic pure states:
\begin{equation}\label{unrave}
\hat\rho_{sys}(t) = M_Z\left[|\psi_Z(t)\rangle\langle\psi_Z(t)|\right].
\end{equation}
Here, $|\psi_Z(t)\rangle$ indicates the solution of the stochastic
Schr\"odinger equation with a particular realization of the 
- in this case Wiener -
stochastic process $Z(t)$, the mean $M_Z[\ldots]$ refers to 
the ensemble average over these processes. The states
$|\psi_Z(t)\rangle$ may or may not be normalized depending on
whether one utilizes the nonlinear \cite{GisPer} or linear 
\cite{GoeGra94} version of the stochastic
Schr\"odinger equation.
Both the linear and the non-linear equation lead to
the correct reduced density operator according to (\ref{unrave})
and they are mathematically equivalent by virtue of a redefinition
of the stochastic processes $Z(t)$ \cite{Ghirar90a}.

It is the aim of this paper to demonstrate
that a stochastic decomposition just like (\ref{unrave}) also holds
in the general case, without any approximation, in particular without
Markov approximation. We derive the linear version of the relevant
non-Markovian stochastic Schr\"odinger equation,
the corresponding non-linear, norm preserving 
theory can be found similar to
the Markovian case \cite{DioStr97c}.

Our result can be based on
two recent independent approaches to a stochastic description 
of non-Markovian open systems \cite{Diosi96,Strunz96b}.
One approach \cite{Diosi96} uses coherent states and has the 
advantage of offering an interpretation of the solutions of the 
stochastic Schr\"odinger equation from first principles.
The other \cite{Strunz96b} is based on the
Feynman-Vernon approach to open systems using path integrals 
\cite{FeyVer63} and is valid for arbitrary temperatures. 

To be specific, we use a standard model of open system quantum mechanics,
a system coupled 
linearly via position coupling to an environment of harmonic oscillators
\cite{FeyVer63}
\begin{equation}\label{sysenv}
\hat H_{tot} = H_{sys}(\hat q,\hat p) - \hat q \sum_i \chi_i \hat Q_i + 
\sum_i \left(\frac{\hat P_i^2}{2m_i} 
+ \frac{1}{2}m_i\omega_i^2 \hat Q_i^2\right),
\end{equation}
where for later purposes we also introduce 
$\hat F = \sum_i \chi_i \hat Q_i$, the 
force acting on the system as induced by the environment.
We assume a factorized total initial density operator
$\hat\rho_{tot}(0) = |\psi_0\rangle \langle\psi_0|
\otimes\hat\rho_{T}.$
The environment oscillators are assumed to be in a thermal initial state 
$\hat\rho_{T}$ at temperature $T$, and, for simplicity, the system 
is assumed to be in a pure state $|\psi_0\rangle\langle\psi_0|$.
The time evolution of the total system is determined by the unitary
von-Neumann equation 
${\dot{\hat \rho}}_{tot} = -i [\hat H_{tot},\hat \rho_{tot}]$.

Without any approximation, the reduced density operator $\hat\rho_{sys}(t)$
of the model (\ref{sysenv}) can be represented as the ensemble average
(\ref{unrave}) of stochastic pure states $|\psi_Z(t)\rangle$. They
are the solutions of the following non-Markovian stochastic 
Schr\"odinger equation
\begin{equation}\label{nonmar}
|\dot{\psi}_Z(t)\rangle = -i H_{sys}(\hat q,\hat p)|\psi_Z(t)\rangle 
 +i \hat q Z(t)|\psi_Z(t)\rangle + i 
\hat q \int_0^t\!\! ds\; \alpha(t,s)\; \frac{\delta |\psi_Z(t)\rangle}
{\delta Z(s)},
\end{equation}
which is the main result of this paper. Equ. (\ref{nonmar}) is a stochastic 
equation, since it depends on a stochastic process
$Z(t)$ specified below. It is also non-Markovian due to a memory
term involving the dependence of the current state $|\psi_Z(t)\rangle$
on earlier noise $Z(s)$, describing the (delayed) back reaction of 
the environment on the system.

The dynamical properties and the temperature of the environment determine 
the memory kernel
$\alpha(t,s)=\sum_i \left(\chi_i^2/2m_i\omega_i\right)\left[\coth
(\hbar\omega_i/2k_BT)\cos\omega_i(t-s)-i\sin(t-s)\right]$ \cite{FeyVer63}. 
It can be regarded as the force correlation function
$\alpha(t,s) = \mbox{Tr}\left(\hat F(t)\hat F(s)\hat \rho_{T}\right),$
where $\hat F(t)$ is the Heisenberg operator of the force of the model 
(\ref{sysenv}) of the undisturbed environment.
This memory kernel also determines the probability distribution of the
stochastic processes $Z(t)$ entering the non-Markovian stochastic
Schr\"odinger equation (\ref{nonmar}). They are
coloured complex Gaussian processes with properties
\begin{equation}\label{colpro}
M_Z[Z(t)] = 0 ,\;\;
M_Z[Z(t)Z(s)] = 0 \;\;\mbox{and}\;\; 
M_Z[Z(t)Z^*(s)] = \alpha^*(t,s),
\end{equation}
designed in such a way as to mimic the effect of the quantum force 
$\hat F(t)$.

In the next two parts we will prove the central assertion of this paper:
the solutions $|\psi_Z(t)\rangle$ of the non-Markovian 
stochastic Schr\"odinger equation (\ref{nonmar}) reproduce the exact
reduced density operator $\hat\rho_{sys}(t)$
if one takes the ensemble average over the stochastic processes $Z(t)$ 
according to (\ref{unrave}).

The first proof uses path integrals.
The propagator $J(t;0)$ of the reduced density 
operator of the model (\ref{sysenv}) can be found in Feynman 
and Vernon's original paper \cite{FeyVer63},
\begin{equation}\label{exapro}
J(q,q',t;q_0,q'_0,0) = \int{\cal D}[q]\int{\cal D}[q'] 
\exp\{ iS_{sys}[q]-iS_{sys}[q']\}\;\; {\cal F}[q,q'],
\end{equation}
with the influence functional ${\cal F}[q,q']$ encoding the effects of
the environment on the system.
It has been shown recently \cite{Strunz96b} that the propagator
(\ref{exapro}) allows an exact stochastic
decomposition using the coloured complex Gaussian stochastic processes $Z(t)$
\cite{foot1} with properties (\ref{colpro}), 
\begin{equation}\label{stodec}
J(q,q',t;q_0,q'_0,0) = M_Z\left[G_Z(q,t;q_0,0) G_Z^*(q',t;q_0',0)\right].
\end{equation}
In \cite{Strunz96b}, the stochastic propagators $G_Z(t;0)$ were given by 
their path integral 
representation
\begin{equation}\label{stopro}
G_Z(q,t;q_0,0) = 
\int{\cal D}[q]\exp\left[ iS_{sys}[q] 
+ i \int_0^t d\tau q_\tau Z(\tau) 
- \int_0^t d\tau \int_0^\tau d\sigma\;
q_\tau\; \alpha(\tau,\sigma)\; q_\sigma\right],
\end{equation}
where we concluded that the states
\begin{equation}\label{stosta}
|\psi_Z(t)\rangle = G_Z(t;0)|\psi_0\rangle
\end{equation}
recover the reduced density operator $\hat\rho_{sys}(t)$ according to 
(\ref{unrave}).

Now we derive the Schr\"odinger equation corresponding to the
stochastic propagator (\ref{stopro}): first we take the time derivative
on both sides of equation (\ref{stosta}). The action $S_{sys}[q]$ in the 
exponent of
the path integral expression (\ref{stopro}) leads to the system
Hamilton operator $\hat H_{sys}$ in the Schr\"odinger equation,
the stochastic integral $\int d\tau q_\tau Z(\tau)$
adds the stochastic driving term $\hat q Z(t)$ in (\ref{nonmar}).
The only complication arises from the contribution of the double time 
integral in the exponent of the action in the path integral expression 
(\ref{stopro}). Using functional differentiation, it is straightforward
to show that it leads to the remaining memory term in (\ref{nonmar}).
This completes the desired proof of the equivalence of the stochastic
propagator (\ref{stopro}) and the non-Markovian stochastic Schr\"odinger
equation (\ref{nonmar}). With (\ref{stodec}) we conclude that
the solutions $|\psi_Z(t)\rangle$ of our
equation (\ref{nonmar}) do in fact recover the
reduced density operator of the model (\ref{sysenv}) by taking the
ensemble average (\ref{unrave}) over the processes $Z(t)$.

It is easy to show - see also \cite{Strunz96b} -
that our equation (\ref{nonmar})
reduces to the well-known (linear) Markovian stochastic Schr\"odinger
equation with complex Wiener noise in the limit of white noise.

An alternative derivation of the non-Markovian 
stochastic Schr\"odinger equation 
(\ref{nonmar}) uses a coherent state basis for the
environmental degrees of freedom. This route to a stochastic
description of non-Markovian open systems was taken in
\cite{Diosi96}. We use the unnormalized Bargmann coherent states 
$|a\rangle$ \cite{KlaSud68} which are analytical in $a$ and
satisfy the completeness relation
$\int d^2a\; e^{-|a|^2}\; 
|a\rangle\langle a | = 1\!\!1,$
where $d^2a \equiv d\mbox{Re}a\, d\mbox{Im}a / \pi$.
Assume an expansion of the total state vector using a
Bargmann basis 
$|a\rangle \equiv |a_1\rangle\otimes|a_2\rangle\otimes\ldots$
for the environmental degrees of freedom $a = (a_1,a_2,\ldots)$,
\begin{equation}\label{expans}
|\Psi_{tot}(t)\rangle =
 \int d^2a\; e^{-|a|^2} \;
|\psi_{a^*}(t)\rangle \otimes |a\rangle.
\end{equation}
The states $|\psi_{a^*}(t)\rangle$ of the system correspond to a particular 
'configuration' $|a\rangle$ of the environment.
Tracing over the environment, using the representation
(\ref{expans}) for the total state, we find that
the reduced density operator takes the form
\begin{equation}\label{unrav3}
\hat \rho_{sys}(t) = \int d^2a\; e^{-|a|^2} \;
|\psi_{a^*}(t)\rangle\langle\psi_{a^*}(t)|
= M_{a}\left[|\psi_{a^*}(t)\rangle
\langle\psi_{a^*}(t)|\right].
\end{equation}
For the last expression we regard the coherent state variables
$a$ as classical stochastic variables with Gaussian 
distribution $M_{a}[\ldots] = \int d^2a[\ldots]\,\exp\{-|a|^2\}$.

We now turn our attention to the time evolution of the total state using
the coherent state representation (\ref{expans}). First,
we rewrite the total
Hamilton operator (\ref{sysenv}) in terms of the creation and annihilation
operators $\hat a^\dagger_i$ and $\hat a_i$ of the environment oscillators, and also
change to a Heisenberg representation of the environmental
part of the total Hamiltonian.
Then we find that the system part
$|\psi_{a^*}(t)\rangle$ of the total state 
obeys the Schr\"odinger equation \cite{KlaSud68}
\begin{equation}\label{cohsch}
|\dot\psi_{a^*}(t)\rangle = -i\hat H_{sys}|\psi_{a^*}(t)\rangle
 + i \hat q \sum_i \frac{\chi_i}{\sqrt{2m_i\omega_i}}\left(
e^{i\omega_i t} a^*_i  + e^{-i\omega_i t}
\frac{\partial}{\partial a^*_i}  
\right)|\psi_{a^*}(t)\rangle.
\end{equation}

We now
restrict ourselves to the zero temperature case ($T=0$), for which
the initial condition $|\psi_{a^*}(0)\rangle = |\psi_0\rangle$
holds for all configurations $a$. In this coherent state approach,
the non-zero temperature case is
non-trivial and will be treated elsewhere \cite{DioStr97c}.

In equ.(\ref{unrav3}) the reduced density operator is expressed naturally
as an ensemble average over the pure states $|\psi_{a^*}(t)\rangle$.
Accordingly, the evolution equation (\ref{cohsch}) represents
the stochastic Schr\"odinger equation for these states.

Remarkably, this construction is identical to the non-Markovian
stochastic Schr\"odinger equation (\ref{nonmar}).
To see the equivalence, we {\it define} stochastic processes
\begin{equation}\label{cohpro}
Z_{a}(t) \equiv \sum_i \frac{\chi_i}{\sqrt{2m_i\omega_i}} a_i^* 
e^{i\omega_i t}.
\end{equation}
A simple calculation shows that these processes are  - for zero temperature -
realizations
of the coloured complex Gaussian stochastic processes (\ref{colpro}):
\begin{equation}\label{coprpr}
M_{a}[Z_{a}(t)] = 0 ,\;\;
M_{a}[Z_{a}(t)Z_{a}(s)] = 0 \;\;\mbox{and}\;\; 
M_{a}[Z_{a}(t)Z_{a}^*(s)] = \alpha_{T=0}^*(t,s).
\end{equation}
Moreover, using the chain rule we find
\begin{equation}\label{dersum}
\sum_i \frac{\chi_i}{\sqrt{2m_i\omega_i}} e^{-i\omega_it} 
\frac{\partial}{\partial a_i^*} = 
\int ds \sum_i\frac{\chi_i^2}{2m_i\omega_i} e^{-i\omega_i(t-s)} 
\frac{\delta}{\delta Z_{a}(s)} =
\int ds \,\alpha_{T=0}(t,s)\,
\frac{\delta}{\delta Z_{a}(s)}.
\end{equation}
Replacing the expressions (\ref{cohpro}) and (\ref{dersum}) in
the coherent state stochastic Schr\"odinger equation (\ref{cohsch}),
we recover the non-Markovian stochastic Schr\"odinger equation 
(\ref{nonmar}), our basic result.

According to (\ref{expans}), the solutions 
$|\psi_Z(t)\rangle$ of the non-Markovian
stochastic Schr\"odinger equation (\ref{nonmar}) represent the part 
of the
total state which corresponds to a certain classical 'configuration' 
$a$ of the environment. The link between the process
$Z(t)$ and the configuration $a$ is given by (\ref{cohpro}).
The interpretation of the classical 'configuration' $a$ of the
environment (and thus of the stochastic processes $Z(t)$ too) can be given 
in the broader framework of hybrid densities \cite{hybrid}.

We have presented 
the non-Markovian stochastic Schr\"odinger equation. It allows
the description of open quantum systems in terms of stochastic state
vectors rather then density operators, without relying on
the Markov approximation.

Two alternative derivations are given. First, we establish the
connection to the path integral approach to open systems as initiated
by Feynman and Vernon. The non-Markovian stochastic Schr\"odinger
equation reflects a stochastic decomposition of the propagator for the
reduced density operator into stochastic propagators for state vectors.
Secondly, we establish the connection to a coherent state description
of the environment, allowing a microscopic interpretation of the
stochastic states $|\psi_Z(t)\rangle$.

Our theory is exact and the model (or some straightforward generalization of it)
appears in many areas of physics
(electron-phonon interaction, the spin-boson model, the whole of quantum 
optics, relativistic field theories as relevant as QED).
These problems can now be phrased in the language of stochastic evolution
equations in Hilbert space, without approximation. The success of quantum 
trajectory methods in the Markovian case suggests that our result 
also represents a promising step towards an effective numerical
algorithm for non-Markovian reduced dynamics.
As a fundamental concept, stochastic pure state representations no longer
depend on the Markov assumption.
Starting from our exact result, approximations like perturbation
expansions are possible, further simplifying the
description of non-Markovian reduced dynamics.

We would like to thank I.C. Percival and J.J. Halliwell for fruitful 
discussions.
W.T.S. thanks the European Union for a Marie Curie fellowship and the
Deutsche Forschungsgemeinschaft for support through the SFB 237 
"Unordnung und gro{\ss}e Fluktuationen".
L.D. was supported by a Visiting Fellowship from EPSRC and by OTKA T016047.


\begin{thebibliography}{999}
\bibitem{foot1} In \cite{Strunz96b} we used $-iZ(t)$ instead of $Z(t)$. Our
theory, however, is invariant under phase changes of the stochastic processes
Z(t), which is why both choices are equivalent.
\end{thebibliography}
\end{document}